\documentclass[twocolumn,nofootinbib,superscriptaddress]{revtex4-1}
\usepackage{latexsym,epsfig,amssymb, amsmath,nicefrac}

\usepackage{color}
\usepackage[makeroom]{cancel}
  \usepackage{latexsym}
  \usepackage{epsf}
  \usepackage{amssymb}
  \usepackage{graphicx}
  \usepackage{amsmath}
  \usepackage{amsmath,amssymb,amsthm}
  \usepackage{verbatim}
  \usepackage{hyperref}

\def\tb0{\tilde{\beta}_0}
{\def\b0{\beta_0}

\def\bi{\begin{itemize}}
\def\ei{\end{itemize}}
\def\be{\begin{equation}}
\def\ee{\end{equation}}
\newcommand{\bea}{\begin{eqnarray}}
\newcommand{\eea}{\end{eqnarray}}

\begin{document}

\vspace{1cm}

\title{The Lyth Bound of Inflation with a Tilt}

\author{Juan Garcia-Bellido}
\email{juan.garciabellido@uam.es}
\affiliation{Instituto de F\'isica Te\'orica IFT-UAM-CSIC, Universidad Aut\'onoma de Madrid,
C/ Nicol\'as Cabrera 13-15, Cantoblanco, 28049 Madrid, Spain}
\author{Diederik Roest}
\email{d.roest@rug.nl}
\author{Marco Scalisi}
\email{m.scalisi@rug.nl}
\author{Ivonne Zavala}
\email{e.i.zavala@rug.nl}
\affiliation{Van Swinderen Institute, University of Groningen, \\ Nijenborgh 4, 9747 AG Groningen, The Netherlands}

\begin{abstract}
We provide strong evidence for universality of the inflationary field range: given an accurate measurement of $(n_s,r)$, one can infer $\Delta \phi$ in a model-independent way in the sub-Planckian regime for a range of universality classes of inflationary models. Both the tensor-to-scalar ratio as well as the spectral tilt are essential for the field range. Given the Planck constraints on $n_s$,  the Lyth bound  is strengthened by two orders of magnitude: whereas the original bound gives a sub-Planckian field range for $r \lesssim 2 \cdot 10^{-3}$, we find that $n_s=0.96$ brings this down to $r \lesssim 2 \cdot 10^{-5}$.

\end{abstract}	

\maketitle

%\smallskip

%\newpage

{\bf Introduction.}\  
Two of the most robust predictions of inflation are a nearly scale invariant spectrum of density perturbations, encoded in the spectral index or tilt $n_s$,  and a stochastic background of gravitational waves, encoded in the tensor-to-scalar ratio $r$. The spectral index has been measured by the Planck satellite \cite{Planck}:
 \begin{align}
  n_s=0.9603\pm0.0073 \,, \label{ns}
 \end{align}
while exact scale invariance  corresponds to  $n_s=1$. Moreover, Planck has placed an upper limit on $r$ of around 10 percent. In contrast, the recent BICEP2 claim \cite{BICEP2} of a detection around 20 percent awaits further clarification and hence will not be considered in this Letter.

A crucial distinction in inflationary models is between small- and large-field models, defined by sub- and super-Planckian field ranges $\Delta \phi$. Generic quantum corrections to a tree-level scalar potential  come in higher powers of $\phi$, and hence large-field models are particularly sensitive to these. This puts the consistency of an effective field theory description of such models into doubt. A key question in theoretical cosmology is therefore whether the inflationary field range exceeds the Planck length or not. 

Knowledge of the evolution of $r(N)$ during all e-foldings $N$ of the inflationary period would determine the field range by means of  $(M_P=1)$
\be \label{phiN}
\frac{d\phi}{d N} = \sqrt{\frac{r(N)}{8}}\,.
\ee
 Moreover, a first estimate of $\Delta \phi$ can be obtained by the assumption that $r(N)$ is constant throughout inflation. This is referred to as the Lyth bound  \cite{Lyth} and leads to \cite{Boubekeur:2005zm, Lotfi}:
\be \label{Lbound}
\Delta\phi \sim \left(\frac{r}{0.002}\right)^{1/2} \left(\frac{N_*}{60}\right) \,,
\ee
where $N_* $ is the number of e-folds at horizon exit, which we  set equal to $60$ (other values allow for a similar analysis). Therefore, a sub-Planckian excursion for the inflaton field requires a very small value of $r\lesssim 2 \cdot 10^{-3}$.

\begin{figure}[t!]
%\vspace*{-5mm}%\hspace{-3mm}
\begin{center}
\includegraphics[width=7.5cm]{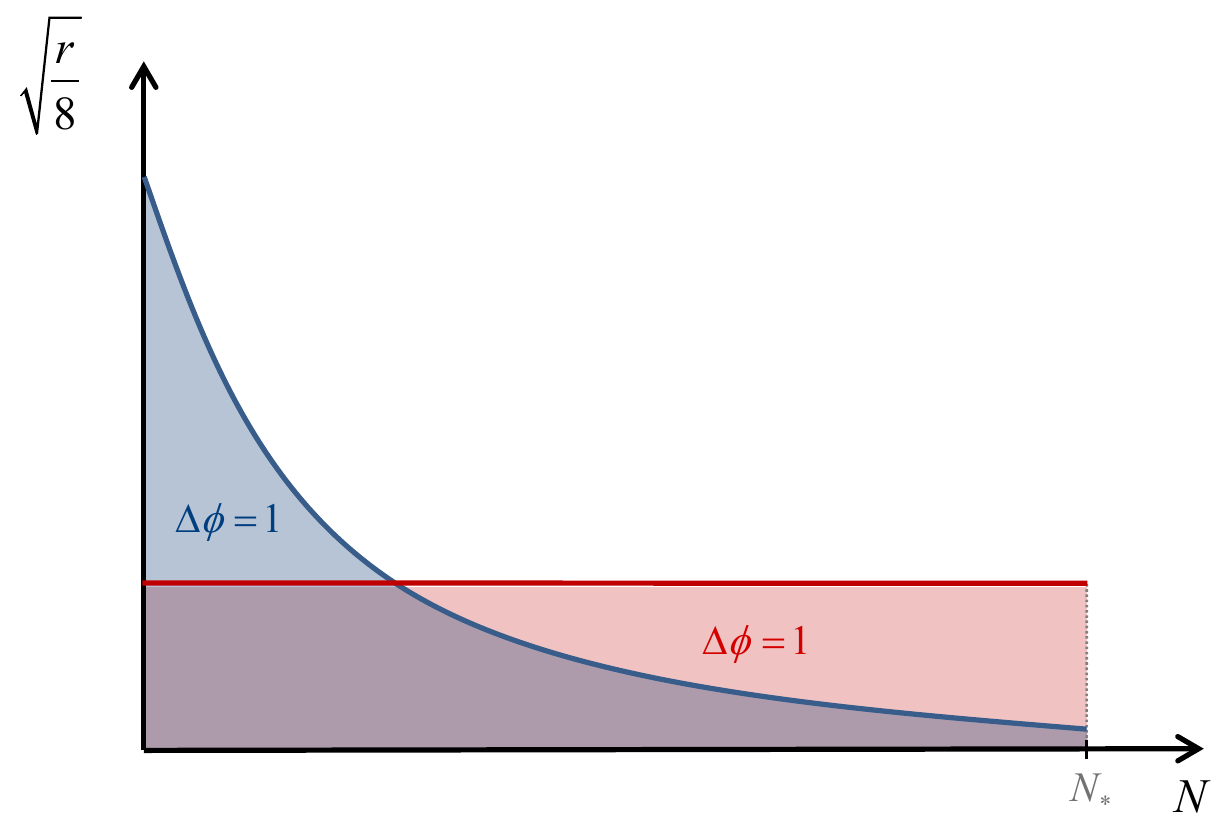}
%\vspace*{-1cm}
\caption{\it Two curves indicating $\sqrt{r(N)/8}$ with identical areas $\Delta \phi=1$. The flat curve depicts the Lyth bound, while the tilted curve indicates the improvement when taking the spectral index into account.}\label{fig1}
\end{center}
\vspace{-0.5cm}
\end{figure}

The Lyth bound provides an optimal estimate of the field range, given a measurement of $r$, which corresponds to  the rectangular area in Fig.~\ref{fig1}. 
However, starting from the same value of $r$ at horizon crossing, one can imagine different behaviors $r(N)$ that give rise to either smaller\footnote{The Lyth bound can also be evaded using multiple scalar \cite{McDonald} or vector fields \cite{MSJ}. An extension   to fast roll can be found in \cite{Baumann}.} \cite{Ido, Hotchkiss, Antusch} or larger areas \cite{GBRSZ}.

We would like to show that this estimate becomes stronger when one takes the additional information of the spectral index into account. In particular, given the redshifted value \eqref{ns} and assuming $r$ to be small, the dependence $r(N)$ is tilted upwards at horizon crossing\footnote{Note that our approach differs from \cite{German:2014qza,Bramante:2014rva}, which also include the spectral tilt in their expressions: while these references derive a {\it minimal} value for $\Delta \phi$, we aim to provide a {\it generic} estimate by making use of its universal properties.}. The natural history therefore leads to a larger area than that of the corresponding rectangle. As a consequence, the requirement  $\Delta \phi=1$ implies a lower value of $r$, as illustrated by the blue line in Fig.~\ref{fig1}. This is our main message: by including constraints on $n_s$ one can strengthen considerably  the Lyth bound. Subject to a number of natural assumptions, after proving universality of $\Delta\phi$ in the sub-Planckian regime,  we will show that the reported value \eqref{ns} leads to $r \lesssim 2 \cdot 10^{-5}$ for sub-Planckian field ranges. This constitutes a bound which is two orders of magnitude stronger than the usual estimate as given by Eq.~\eqref{Lbound}.
 \\

{\bf Universality at large $N$.}\ The Planck reported value \eqref{ns}  is consistent with the simple Ansatz of a tilt $n_s$ whose deviation from unity scales with $1/N$. In fact, for around 60 e-folds, this gives a percent-level deviation from scale invariance. This assumption leads naturally to a first slow-roll parameter $\epsilon(N)=r(N)/16$ that scales as a power of $1/N$ \cite{Mukhanov, Roest, GarRoe} (see also \cite{Huang}):
\be \label{epsP}
\epsilon = \frac{\beta}{N^p}\,,
\ee
with $\beta$ and $p$ being constant. This simple Ansatz leads to
\begin{align} \label{puniv}
   r =  \frac{16\beta}{N^p} \,, \qquad
   n_s=\begin{cases}
1 - \frac{2\beta+1}{N} \,, \quad &p=1\,, \\
1 - \frac{p}{N} \,, \quad &p>1\,, 
 \end{cases}
\end{align}
where the case $p<1$ has been discarded as it generically leads to values of the cosmological observables not compatible with the current data. Eq.~\eqref{puniv} identifies the families of universality classes which any specific scenario belongs to, for fixed values of $\beta$ and $p$. Subleading corrections have  higher powers in $1/N$ and are irrelevant from the observational viewpoint. Several examples are listed in \cite{Roest, GarRoe}. 

In a pure large-$N$ description, one can identify the benchmark potentials for this Ansatz. Let us recall that  $\epsilon$ is related to the Hubble parameter $H$ through
\be\label{epsH}
\epsilon= \frac{d \ln H}{d N}\,.
\ee
Within the slow-roll approximation, employing $H^2=V$, one can integrate Eq.~\eqref{epsH} and obtain an expression for the potential in terms of $N$ which reads
\begin{align} \label{VN}
V(N)=\begin{cases}
V_0\,N^{2\beta} \,, \quad &p=1\,, \\
V_0\left[1 - \frac{2\beta}{(p-1) N^{p-1}}\right] \,, \quad &p>1\,,
 \end{cases}
 \end{align}
where $V_0$ is an integration constant related to the energy scale of inflation.  By means of Eq. \eqref{phiN} and \eqref{epsP}, one  gets the asymptotic form of $V$ in terms of the canonical scalar field $\phi$, that is
\bea \label{VP}
V(\phi)=\begin{cases}
V_0\,\phi^n \,, \quad &p=1\,, \\
V_0\left[1-\exp\left(- \phi/\mu\right)\right] \,, \quad &p=2\,, \\
V_0\left[1-\left(\phi/\mu\right)^{n}\right] \,, \quad &p>1\,, p\neq2\,,
 \end{cases}
 \eea
where $\mu$ and $n$ are related to $\beta$ and $p$ as dictated by \eqref{phiN}. In particular, for $p>1$ and $p\neq2$, the power  $n$ is related to $p$ through the following equation
\be\label{np}
n=\frac{2(1-p)}{2-p}\,,
\ee 
where $p<2$ or $p>2$  determine respectively the negative or positive sign of $n$. The inverse relation $p=p(n)$ turns out to be of the same form. 

In the large-$N$ limit, any model belonging to these universality classes will have a potential asymptotically approaching well-known scenarios such as chaotic monomial inflation ($p=1$), inverse-hilltop models ($1<p<2$), Starobinsky-like inflation ($p=2$) and hilltop potentials ($p>2$). The reason for such  simplicity is that, in this limit, we are probing just a limited part of the inflationary trajectory,  close to  horizon crossing. Peculiarities among different models appear  when we go away from this region. In general, the situation near the end-point of inflation will be very different from one model to another, even though they belong to the same universality class. 

Following the above reasoning, one would expect that a variable such as the inflaton excursion $\Delta\phi$, which evidently depends on the entire inflationary trajectory, does not manifest any universality feature. Nevertheless, it is possible to identify different regions where the field range does exhibit a universal behavior.

In order to get the expression for $\Delta\phi$, one must integrate Eq.~\eqref{phiN} along the entire inflationary trajectory. By considering a large-$N$ behavior such as that in Eq.~\eqref{epsP}, for $p\neq2$, we obtain
\be \label{range}
\Delta\phi= \frac{2\sqrt{2\beta}}{2-p} \,N^{1-\frac{p}{2}}-\phi_e\,,
\ee
where $\phi_e$ is a constant piece related to the value of the inflaton when inflation ends. Then, we run into two possible  situations, depending on whether $p$ is smaller or larger than $2$.

In the first case, for $p<2$, the inflaton range $\Delta \phi$  is  proportional to a positive power of $N$. In the large-$N$ limit, the constant part $\phi_e$ is subleading and one can  argue that, within any universality class, the magnitude of field excursion will be model-independent and therefore universal. Furthermore, given that $\Delta \phi$ keeps increasing together with $N$, one can correctly refer to such scenarios as genuine large field models. 

In the second case, for $p>2$, the $N$-dependent term of \eqref{range} is subleading with respect the constant term $\phi_e$, in the large-$N$ limit. The value of $\Delta \phi$ is therefore determined by the point where inflation stops and generically not universal: for instance, $\Delta\phi$ can already obtain  a super-Planckian contribution during the last e-fold \cite{EKP}. This model-dependent piece is generically sub-dominant for models with $p<2$ while it represents the main contribution when $p>2$.  

Finally, the remaining possibility is $p=2$ where the functional form of the field range reads
\be \label{range2}
\Delta\phi= \sqrt{2\beta}\, \ln N-\phi_e\,.
\ee

The log-dependence leads to a situation where $\Delta \phi$ mildly increases together with $N$. The special role of this point, corresponding to Starobinky-like scenarios, has been recently highlighted in the context of the inflationary attractors \cite{Kallosh:2013tua, Kallosh:2013yoa} as well as non-compact symmetry breaking \cite{Burgess}. Moreover, a change of behavior around the point $p=2$ was noticed also in the analysis on the degeneracy of the inflaton range done in \cite{GBRSZ}.  Here we stress its peculiarity  also as  marking the separation between a region of authentic large field models ($p<2$), whose $\Delta \phi$ exhibits universality features, and a region ($p>2$) where models can have the same $r$ and $n_s$ at leading order (and, thus, belonging to the same universality class) but still very different field ranges. 
\\

{\bf Universality at small $\mu$.}\ The results presented above are obtained in a pure large-$N$ expansion, that is, in the limit $N\rightarrow \infty$. However, physical values usually amount to an exponential expansion of around 50 to 60 e-foldings preceding the end of inflation. Although the latter is a big number, the universal regime can be easily affected by tuning specific parameters of the models.

For large enough values of $N$, any model, characterized by an equation of state parameter such as Eq.~\eqref{epsP}, will be represented by a potential parameterized as a small deviation from the benchmarks \eqref{VP}. Specifically, for $p>1$ and $p\neq 2$, the generic form of $V$ will include higher order corrections and read
\be\label{Vdev}
V(\phi) = V_0 \left[1-\left(\frac{\phi}{\mu}\right)^{n} + \sum^{\pm\infty}_{q=n\pm 1} c_q \left(\frac{\phi}{\mu}\right)^q\right]\,,
\ee 
where $n$ is related to $p$ through Eq.~\eqref{np} and the plus or minus sign depends respectively on $p>2$ or $p<2$. Then, the coefficients $c_q$ parameterize the deviation  from hilltop or inverse-hilltop models respectively.

We now show that, at small $\mu$ and for finite values of $N$, we recover universality: in addition to the cosmological observables $n_s$ and $r$, the inflaton excursion will be model-independent. Interestingly, this is exactly the regime we will consider  to derive  the {\it field range bound}.

The spectral index $n_s$ and tensor-to-scalar ratio $r$ will be generically insensitive to higher order terms in the expansion \eqref{Vdev} as they are calculated at horizon exit. In fact, the inflationary regime is restricted to the region $\phi<\mu$, for hilltop models ($p>2$), and $\phi>\mu$, for inverse hilltop potentials ($1<p<2$); therefore, the farther one is located from the end-point of inflation the more one can ignore higher order corrections in the scalar potential. Then, the large-$N$ regime provides an accurate estimate of such observables which, at small $\mu$, read
\bea \label{nsrHT}
n_s&=& 1-\frac{p}{N} \,,\quad 
r = 2^{5-2p} \frac{(p-2)^{2p-2}}{(p-1)^{p-2}}\frac{\mu^{2p-2}}{N^p}\,.
\eea
The coefficients $c_q$ will appear only in subleading terms in $N$. The family of models represented by Eq.~\eqref{Vdev} will have identical behavior in the small-$\mu$ limit and for large enough values of $N$. Conversely, this is generically not the case for large values of $\mu$; in such a limit, the end-point of inflation is pushed  towards the region where the coefficients $c_q$ play an important role and dissimilarities become important; consequently, going 50-60 e-foldings back, even the point at horizon crossing will start to be sensitive to $c_q$ corrections. For large values of $\mu$, the large-$N$ expansion is not well defined and scenarios belonging to the same universality class at small $\mu$, may give quite different predictions in terms of $n_s$ and $r$. 

In the limit of large $N$ and small $\mu$, the field range turns out to be
\be\label{deltaVdev}
\begin{aligned}
\Delta\phi= & \left[\frac{2-p}{\sqrt{2}(1-p)}\right]^{1-\frac{2}{p}}\mu^{2-\frac{2}{p}} - \frac{(\frac{p}{2}-1)^{p-2}}{(p-1)^{\frac{p}{2}-1}}\, \mu^{p-1} N^{1-\frac{p}{2}}\,,
\end{aligned}
\ee
where the first term is clearly related to the end-point of inflation while the second one is the $N$-dependent term. For the  reasons given above, $c_q$ corrections will not enter the $N$-dependent part which gives the main contribution to the field range for $1<p<2$ while it is subleading for $p>2$. Things are different when calculating the end-point $\phi_e$;  this piece is sensitive to higher-order corrections in $\mu$. As soon as $\mu$ increases, this point is pushed away towards a region where differences among the models begin to appear. If, for simplicity, we focus on the case $n=3$ (examples belonging to this universality class are hilltop inflation and the models referred to as RIPI and MSSMI in \cite{MRV}) and consider terms up to fifth order in the expansion \eqref{Vdev}, the end-point reads
\be\label{phie3}
\phi_e= \sqrt{\frac{\sqrt{2}}{3}}\, \mu^{3/2}+ \frac{2\sqrt{2}}{9}\,c_4 \mu^2 + \frac{5(4\,c_4^2 + 3 \,c_5)}{27\cdot 2^{1/4} \sqrt{3}}\mu^{5/2}\,.
\ee 
Crucially, the coefficients $c_q$ appear just with higher powers of $\mu$; this holds even for other values of $n$ (both positive and negative) as well as the special point $p=2$. This implies that one obtains universal predictions in the small-$\mu$ limit, not just in terms of $n_s$ and $r$, but also in terms of $\Delta\phi$, whose form approaches Eq.~\eqref{deltaVdev}.
\\

{\bf Strengthening the Lyth bound. }\ We now use the results derived above in order to revisit the discussion on small- and large-field excursions and derive a stronger field range bound than the usual estimate Eq.~\eqref{Lbound}.

The findings on the universality of the field range translate into the possibility of inferring an accurate estimate of $\Delta\phi$ given a point in the $(n_s,r)$ plane. This is certainly true in the small-$\mu$ limit where $\Delta\phi$ is given by Eq.~\eqref{deltaVdev}. One can properly argue that sub-Planckian field ranges will be model-independent and uniquely determined by a measurement of cosmological observables. The situation changes when $\mu$ increases; already for $\mu \gtrsim \mathcal{O}(1)$, in the region $p>2$ (corresponding to $n_s\lesssim0.96$), universality breaks down (as can be seen from Eq.~\eqref{phie3} where each contribution is order one); differently, for $p<2$, universality can hold even for some orders of magnitude larger than the reduced Planck mass $M_P=1$, thanks to the dominant $N$-dependent term as set by Eq.~\eqref{range}.

\begin{figure}[t!]
\begin{center}
\includegraphics[width=8.5cm]{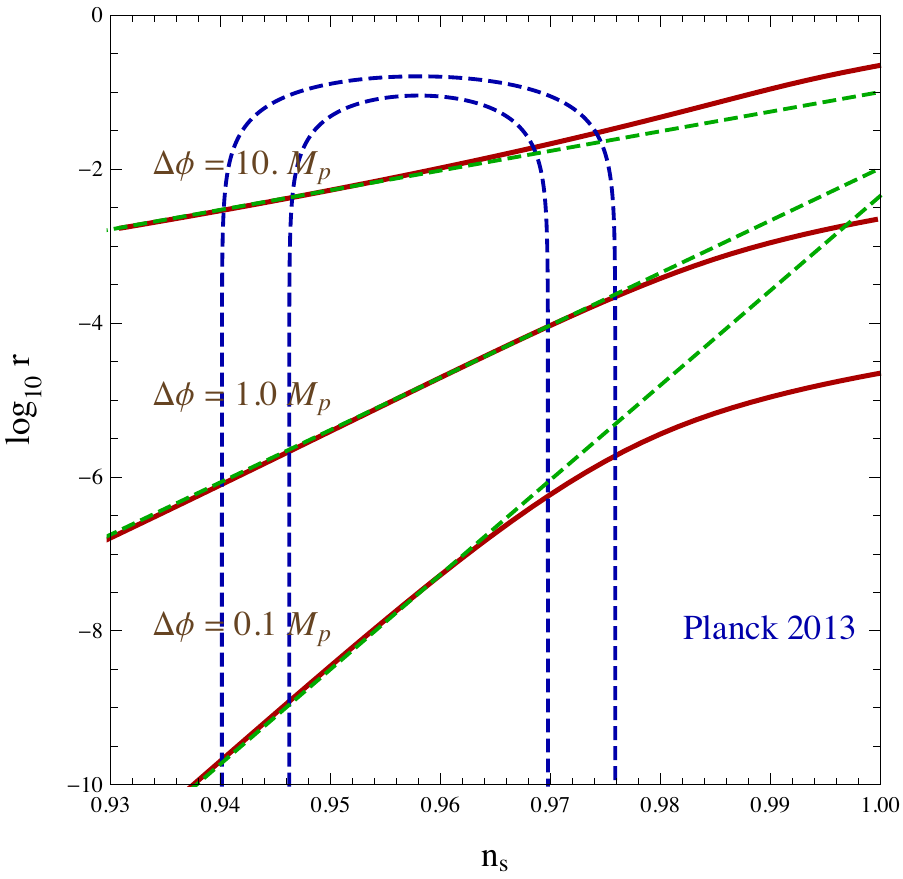}
%\vspace*{-1cm}
\caption{\it\it  Field ranges corresponding to $\Delta \phi = (0.1,1,10)$ in the plane ($n_s,~\log_{10}{(r)}$). The green straight dashed lines represent the asymptotic behaviour for large $p$.} \label{fig.LythBound}
\end{center}
\vspace{-0.5cm}
\end{figure}

Then, if we plot lines of constant $\Delta\phi$ in a  $(n_s,r)$ plane, the one corresponding to unity $\Delta\phi=1$ will be a good estimate of the border above which universality breaks down, regardless the value of $n_s$. This will be taken as the new, stronger bound. As can be seen from Fig.~\ref{fig.LythBound}, the line is tilted as it is a function also of the spectral index $n_s$. Interestingly, for $n_s=1$ it approaches the value of the original Lyth bound, which is a constant value not depending on the tilt. On the other hand, in the Planck-range, an excellent fit is provided by the following expressions, corresponding to the (green) dashed straight lines in Fig.~\ref{fig.LythBound},

\be\label{improved-Lyth}
\begin{aligned}
   \log_{10}r  & = & -1.0 + 25.5\,(n_s -1)\,, \hspace{9mm} \Delta\phi = 10 \,,\\
   \log_{10}r  & = & -2.0 + 68.0\,(n_s -1)\,, \hspace{8mm} \Delta\phi = 1.0 \,,\\
   \log_{10}r  & = & -2.35 + 123\,(n_s -1)\,, \hspace{8mm} \Delta\phi = 0.1 \,. 
 \end{aligned}
\ee

The range of values of ($n_s,\,r$) consistent within those of Planck2013 reduces the values of $\Delta\phi$ during inflation by at least an order of magnitude. For the central value $n_s\simeq 0.96$, imposing that $\Delta\phi \leq 1$ leads to the bound $r\lesssim 2\cdot10^{-5}$, which is two orders of magnitude below the usual Lyth bound.

\begin{figure}[h!]
\vspace*{5mm}
\begin{center}
\includegraphics[width=8.5cm]{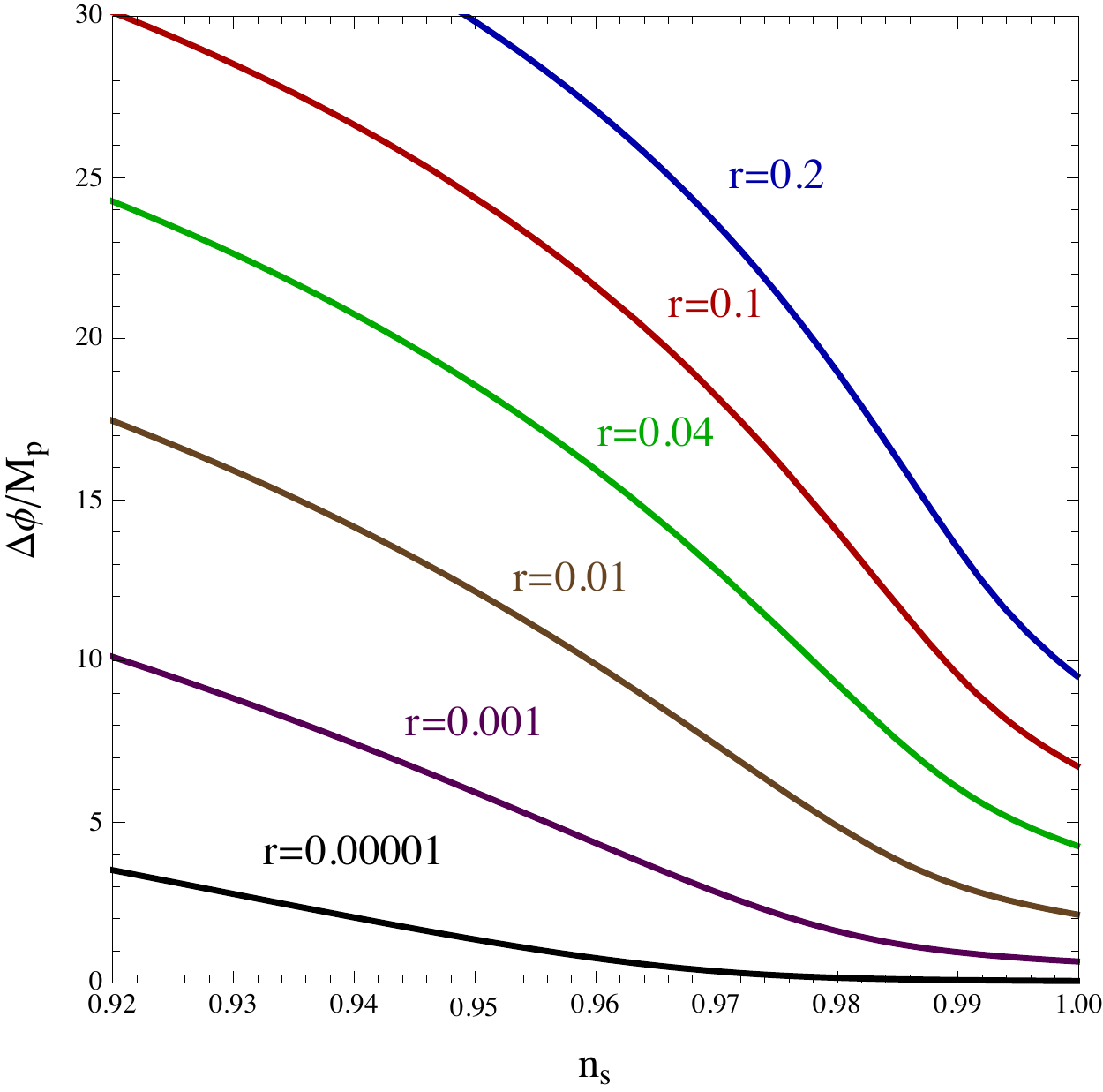}
%\vspace*{-1cm}
\caption{\it The range of field values corresponding to $r = 0.2,\,0.1,\,004,\,0.01,\,0.001,\,0.00001$ in the plane ($n_s,~\Delta\phi$).}\label{fig.dphins}
\end{center}
\vspace{-0.5cm}
\end{figure}

On the other hand, if we impose that the ratio $r$ be bigger than a certain value, then we find a lower bound on $\Delta\phi$. Fig.~\ref{fig.dphins} shows the field range as a function of the scalar spectral index for different values of the ratio $r$. Again, in the range consistent with Planck2013, the field range is always super-Planckian, for all values of the ratio $r \gtrsim 2 \cdot 10^{-5}$. This conclusion can only be avoided by going to unrealistically large spectral indices $n_s$ close to $1$. 
\\

{\bf Discussion.}\ The main results of this Letter are twofold. First of all, we have provided strong arguments for the universality of small field ranges
\footnote{Strictly speaking, this is true for values $\Delta\phi \lesssim 10^{-2}$, which define more accurately  small field inflation. In this  region $\mu<1$ and thus sub-leading corrections are suppressed, strengthening the results on universality.}  $\Delta \phi<1$ as given in \eqref{deltaVdev}.
Secondly, we have pointed out that this results in a significant strengthening on the Lyth bound when including both the spectral index and the tensor-to-scalar ratio, see \eqref{improved-Lyth} and Fig.~\ref{fig.LythBound}.

Similarly to the original Lyth bound, the relations \eqref{improved-Lyth} provide generic estimates of the field range, which could be avoided only by a very specific (non-generic) behavior of $\epsilon(N)$. However the existence of such counterexamples is of limited importance: one would like to understand when large field inflation is expected given a measurement of $r$ even if there might be fine-tuned models which give smaller field ranges for this value of~$r$. 

Given the central value for $n_s$ from Planck, our results imply that super-Planckian field ranges require a tensor-to-scalar ratio that exceeds $2 \cdot 10^{-5}$. Planned future CMB experiments, such as COrE \cite{Bouchet:2011ck, Core} and PRISM \cite{Andre:2013afa,Andre:2013nfa, Prism}, might bring the sensitivity down to $10^{-4}$. In contrast to what one would conclude from the original Lyth bound, our results  imply that a small detectable $r$ still corresponds to super-Planckian field ranges.  
\\

{\bf Acknowledgments.}\ We would like to thank David Lyth and David Nolde for useful discussions and Gianmassimo Tasinato for comments on the manuscript. We acknowledge financial support from the Madrid Regional Government (CAM) under the program HEPHACOS S2009/ESP-1473-02, from the Spanish MINECO under grant FPA2012-39684-C03-02 and Consolider-Ingenio 2010 PAU (CSD2007-00060), from the Centro de Excelencia Severo Ochoa Programme, under grant SEV-2012-0249, as well as from the European Union Marie Curie Initial Training Network UNILHC PITN-GA-2009-237920.

\bibliography{refsLyth}

\providecommand{\href}[2]{#2}\begingroup\raggedright\begin{thebibliography}{10}

\bibitem{Planck}
{\bf Planck Collaboration} Collaboration, P.~Ade {\em et al.}, ``{Planck 2013
  results. XXII. Constraints on inflation}'',
\href{http://arxiv.org/abs/1303.5082}{{\tt arXiv:1303.5082 [astro-ph.CO]}}.
%%CITATION = ARXIV:1303.5082;%%.

\bibitem{BICEP2}
{\bf BICEP2 Collaboration} Collaboration, P.~Ade {\em et al.}, ``{Detection of
  B-Mode Polarization at Degree Angular Scales by BICEP2}'',
  \href{http://dx.doi.org/10.1103/PhysRevLett.112.241101}{{\em Phys.Rev.Lett.}
  {\bf 112} (2014)  241101},
\href{http://arxiv.org/abs/1403.3985}{{\tt arXiv:1403.3985 [astro-ph.CO]}}.
%%CITATION = ARXIV:1403.3985;%%.

\bibitem{Lyth}
D.~H. Lyth, ``{What would we learn by detecting a gravitational wave signal in
  the cosmic microwave background anisotropy?}'',
  \href{http://dx.doi.org/10.1103/PhysRevLett.78.1861}{{\em Phys.Rev.Lett.}
  {\bf 78} (1997)  1861--1863},
\href{http://arxiv.org/abs/hep-ph/9606387}{{\tt arXiv:hep-ph/9606387
  [hep-ph]}}.
%%CITATION = HEP-PH/9606387;%%.

\bibitem{Boubekeur:2005zm}
L.~Boubekeur and D.~Lyth, ``{Hilltop inflation}'',
  \href{http://dx.doi.org/10.1088/1475-7516/2005/07/010}{{\em JCAP} {\bf 0507}
  (2005)  010},
\href{http://arxiv.org/abs/hep-ph/0502047}{{\tt arXiv:hep-ph/0502047
  [hep-ph]}}.
%%CITATION = HEP-PH/0502047;%%.

\bibitem{Lotfi}
L.~Boubekeur, ``{Theoretical bounds on the tensor-to-scalar ratio in the cosmic
  microwave background}'',
  \href{http://dx.doi.org/10.1103/PhysRevD.87.061301}{{\em Phys.Rev.} {\bf D87}
  (2013) no.~6, 061301},
\href{http://arxiv.org/abs/1208.0210}{{\tt arXiv:1208.0210 [astro-ph.CO]}}.
%%CITATION = ARXIV:1208.0210;%%.

\bibitem{McDonald}
J.~McDonald, ``{Sub-Planckian Two-Field Inflation Consistent with the Lyth
  Bound}'', \href{http://dx.doi.org/10.1088/1475-7516/2014/09/027}{{\em JCAP}
  {\bf 1409} (2014) no.~09, 027},
\href{http://arxiv.org/abs/1404.4620}{{\tt arXiv:1404.4620 [hep-ph]}}.
%%CITATION = ARXIV:1404.4620;%%.

\bibitem{MSJ}
A.~Maleknejad and M.~Sheikh-Jabbari, ``{Gauge-flation: Inflation From
  Non-Abelian Gauge Fields}'',
  \href{http://dx.doi.org/10.1016/j.physletb.2013.05.001}{{\em Phys.Lett.} {\bf
  B723} (2013)  224--228},
\href{http://arxiv.org/abs/1102.1513}{{\tt arXiv:1102.1513 [hep-ph]}}.
%%CITATION = ARXIV:1102.1513;%%.

\bibitem{Baumann}
D.~Baumann and D.~Green, ``{A Field Range Bound for General Single-Field
  Inflation}'', \href{http://dx.doi.org/10.1088/1475-7516/2012/05/017}{{\em
  JCAP} {\bf 1205} (2012)  017},
\href{http://arxiv.org/abs/1111.3040}{{\tt arXiv:1111.3040 [hep-th]}}.
%%CITATION = ARXIV:1111.3040;%%.

\bibitem{Ido}
I.~Ben-Dayan and R.~Brustein, ``{Cosmic Microwave Background Observables of
  Small Field Models of Inflation}'',
  \href{http://dx.doi.org/10.1088/1475-7516/2010/09/007}{{\em JCAP} {\bf 1009}
  (2010)  007},
\href{http://arxiv.org/abs/0907.2384}{{\tt arXiv:0907.2384 [astro-ph.CO]}}.
%%CITATION = ARXIV:0907.2384;%%.

\bibitem{Hotchkiss}
S.~Hotchkiss, A.~Mazumdar, and S.~Nadathur, ``{Observable gravitational waves
  from inflation with small field excursions}'',
  \href{http://dx.doi.org/10.1088/1475-7516/2012/02/008}{{\em JCAP} {\bf 1202}
  (2012)  008},
\href{http://arxiv.org/abs/1110.5389}{{\tt arXiv:1110.5389 [astro-ph.CO]}}.
%%CITATION = ARXIV:1110.5389;%%.

\bibitem{Antusch}
S.~Antusch and D.~Nolde, ``{BICEP2 implications for single-field slow-roll
  inflation revisited}'',
  \href{http://dx.doi.org/10.1088/1475-7516/2014/05/035}{{\em JCAP} {\bf 1405}
  (2014)  035},
\href{http://arxiv.org/abs/1404.1821}{{\tt arXiv:1404.1821 [hep-ph]}}.
%%CITATION = ARXIV:1404.1821;%%.

\bibitem{GBRSZ}
J.~Garcia-Bellido, D.~Roest, M.~Scalisi, and I.~Zavala, ``{Can CMB data
  constrain the inflationary field range?}'',
  \href{http://dx.doi.org/10.1088/1475-7516/2014/09/006}{{\em JCAP} {\bf 1409}
  (2014)  006},
\href{http://arxiv.org/abs/1405.7399}{{\tt arXiv:1405.7399 [hep-th]}}.
%%CITATION = ARXIV:1405.7399;%%.

\bibitem{German:2014qza}
G.~German, ``{On the Lyth bound and single field slow-roll inflation}'',
\href{http://arxiv.org/abs/1405.3246}{{\tt arXiv:1405.3246 [astro-ph.CO]}}.
%%CITATION = ARXIV:1405.3246;%%.

\bibitem{Bramante:2014rva}
J.~Bramante, S.~Downes, L.~Lehman, and A.~Martin, ``{Clearing the Brush: The
  Last Stand of Solo Small Field Inflation}'',
  \href{http://dx.doi.org/10.1103/PhysRevD.90.023530}{{\em Phys.Rev.} {\bf D90}
  (2014)  023530},
\href{http://arxiv.org/abs/1405.7563}{{\tt arXiv:1405.7563 [astro-ph.CO]}}.
%%CITATION = ARXIV:1405.7563;%%.

\bibitem{Mukhanov}
V.~Mukhanov, ``{Quantum Cosmological Perturbations: Predictions and
  Observations}'', \href{http://dx.doi.org/10.1140/epjc/s10052-013-2486-7}{{\em
  Eur.Phys.J.} {\bf C73} (2013)  2486},
\href{http://arxiv.org/abs/1303.3925}{{\tt arXiv:1303.3925 [astro-ph.CO]}}.
%%CITATION = ARXIV:1303.3925;%%.

\bibitem{Roest}
D.~Roest, ``{Universality classes of inflation}'',
  \href{http://dx.doi.org/10.1088/1475-7516/2014/01/007}{{\em JCAP} {\bf 01}
  (2014)  007},
\href{http://arxiv.org/abs/1309.1285}{{\tt arXiv:1309.1285 [hep-th]}}.
%%CITATION = ARXIV:1309.1285;%%.

\bibitem{GarRoe}
J.~Garcia-Bellido and D.~Roest, ``{The large-N running of the spectral index of
  inflation}'', \href{http://dx.doi.org/10.1103/PhysRevD.89.103527}{{\em
  Phys.Rev.} {\bf D89} (2014)  103527},
\href{http://arxiv.org/abs/1402.2059}{{\tt arXiv:1402.2059 [astro-ph.CO]}}.
%%CITATION = ARXIV:1402.2059;%%.

\bibitem{Huang}
Q.-G. Huang, ``{Constraints on the spectral index for the inflation models in
  string landscape}'', \href{http://dx.doi.org/10.1103/PhysRevD.76.061303}{{\em
  Phys.Rev.} {\bf D76} (2007)  061303},
\href{http://arxiv.org/abs/0706.2215}{{\tt arXiv:0706.2215 [hep-th]}}.
%%CITATION = ARXIV:0706.2215;%%.

\bibitem{EKP}
R.~Easther, W.~H. Kinney, and B.~A. Powell, ``{The Lyth bound and the end of
  inflation}'', \href{http://dx.doi.org/10.1088/1475-7516/2006/08/004}{{\em
  JCAP} {\bf 0608} (2006)  004},
\href{http://arxiv.org/abs/astro-ph/0601276}{{\tt arXiv:astro-ph/0601276
  [astro-ph]}}.
%%CITATION = ASTRO-PH/0601276;%%.

\bibitem{Kallosh:2013tua}
R.~Kallosh, A.~Linde, and D.~Roest, ``{A universal attractor for inflation at
  strong coupling}'',
  \href{http://dx.doi.org/10.1103/PhysRevLett.112.011303}{{\em Phys.Rev.Lett.}
  {\bf 112} (2014)  011303},
\href{http://arxiv.org/abs/1310.3950}{{\tt arXiv:1310.3950 [hep-th]}}.
%%CITATION = ARXIV:1310.3950;%%.

\bibitem{Kallosh:2013yoa}
R.~Kallosh, A.~Linde, and D.~Roest, ``{Superconformal Inflationary
  $\alpha$-Attractors}'', \href{http://dx.doi.org/10.1007/JHEP11(2013)198}{{\em
  JHEP} {\bf 1311} (2013)  198},
\href{http://arxiv.org/abs/1311.0472}{{\tt arXiv:1311.0472 [hep-th]}}.
%%CITATION = ARXIV:1311.0472;%%.

\bibitem{Burgess}
C.~Burgess, M.~Cicoli, F.~Quevedo, and M.~Williams, ``{Inflating with Large
  Effective Fields}'',
\href{http://arxiv.org/abs/1404.6236}{{\tt arXiv:1404.6236 [hep-th]}}.
%%CITATION = ARXIV:1404.6236;%%.

\bibitem{MRV}
J.~Martin, C.~Ringeval, and V.~Vennin, ``{Encyclopaedia Inflationaris}'',
\href{http://arxiv.org/abs/1303.3787}{{\tt arXiv:1303.3787 [astro-ph.CO]}}.
%%CITATION = ARXIV:1303.3787;%%.

\bibitem{Bouchet:2011ck}
{\bf COrE Collaboration} Collaboration, F.~Bouchet {\em et al.}, ``{COrE
  (Cosmic Origins Explorer) A White Paper}'',
\href{http://arxiv.org/abs/1102.2181}{{\tt arXiv:1102.2181 [astro-ph.CO]}}.
%%CITATION = ARXIV:1102.2181;%%.

\bibitem{Core}
\url{http://www.core-mission.org}.

\bibitem{Andre:2013afa}
{\bf PRISM Collaboration} Collaboration, P.~Andre {\em et al.}, ``{PRISM
  (Polarized Radiation Imaging and Spectroscopy Mission): A White Paper on the
  Ultimate Polarimetric Spectro-Imaging of the Microwave and Far-Infrared
  Sky}'',
\href{http://arxiv.org/abs/1306.2259}{{\tt arXiv:1306.2259 [astro-ph.CO]}}.
%%CITATION = ARXIV:1306.2259;%%.

\bibitem{Andre:2013nfa}
{\bf PRISM Collaboration} Collaboration, P.~André {\em et al.}, ``{PRISM
  (Polarized Radiation Imaging and Spectroscopy Mission): An Extended White
  Paper}'', \href{http://dx.doi.org/10.1088/1475-7516/2014/02/006}{{\em JCAP}
  {\bf 1402} (2014)  006},
\href{http://arxiv.org/abs/1310.1554}{{\tt arXiv:1310.1554 [astro-ph.CO]}}.
%%CITATION = ARXIV:1310.1554;%%.

\bibitem{Prism}
\url{http://www.prism-mission.org}.

\end{thebibliography}\endgroup
\bibliographystyle{utphys}

\end{document}